\documentclass[fleqn,twoside]{article}
\usepackage{amsmath}
\usepackage{espcrc2}

\begin{document}

\title{Neutrino Condensate as Origin of Dark Energy}
%

\author{Jitesh R. Bhatt$^a$, Bipin R. Desai$^{b}$, Ernest Ma$^{b}_{}$,
G. Rajasekaran$^{c,d}_{}$ and Utpal Sarkar$^a$\\[.24in] 
$^{\rm a}$ Physical Research Laboratory, Ahmedabad 380 009, India\\
$^{\rm b}$ Physics Department, University of California,
Riverside, CA 92521, USA \\
$^{\rm c}_{}$ Institute of Mathematical Sciences, Chennai 600 113,
India \\
${\rm d}_{}$Chennai Mathematical Institute, Siruseri, 603103, India \\[0.2in]}


\begin{abstract}%
We propose a new solution to the origin of dark energy. We suggest that it
was created dynamically from the condensate of a singlet neutrino at a late
epoch of the early Universe through its effective self interaction. This
singlet neutrino is also the Dirac partner of one of the three observed
neutrinos, hence dark energy is related to neutrino mass. The onset of this
condensate formation in the early Universe is also related to matter density
and offers an explanation of the coincidence problem of why dark energy
(70\%) and total matter (30\%) are comparable at the present time. We
demonstrate this idea in a model of neutrino mass with (right-handed)
singlet neutrinos and a singlet scalar.
\vspace{1pc}%
\end{abstract}%
\maketitle%

\bigskip 

The astrophysical observations that the baryonic and dark matter together
account for only about 30\% of the total matter while the remaining 70\%
have the property of producing negative pressure, interpreted as the dark
energy, remains a challenging problem in cosmology. A possible explanation
that the dark energy is simply the vacuum energy given by the standard model
scale fails by many orders of magnitude. It also can not explain the fact
that the dark energy has been comparable to the density of ordinary matter
during the recent epoch in the evolution of the universe. While solutions to
this problem have been provided, notably through the presence of a scalar
field called quintessence[1],[2]the observation that the magnitude of the
dark energy is comparable to neutrino masses suggests to us that the
explanation of the dark energy puzzle may possibly lie in the direction of
neutrinos,while the other large effects coming from electroweak breaking and
QCD are canceled out by some unknown dynamical effects It is this avenue
which we follow in our discussion below.

Neutrinos have been invoked in the past in the dark energy problems. For
example, recently it has been argued in a theory with scalar fields, called
accelerons, that the dark energy can be obtained dynamically if the neutrino
masses are allowed to vary with the scalar field [3],[4]. We propose,
however, a somewhat more direct connection through the formation of neutrino
condensates [5],[6],[7],[8]. We elaborate below

Our starting point is an extension of the standard model which includes
right-handed neutrinos $\nu _{iR},~i=1,2,3$, in which the neutrino masses
have the seesaw structure [9]with both the Dirac and Majorana masses of the
order of eV. In this model we introduce a light real singlet scalar $S$, so
that the effective Lagrangian after the electroweak symmetry breaking is
given by 
\begin{equation}
\mathcal{L}=m_{Di\alpha }\bar{\ell}_{i}\nu _{\alpha R}+M_{\alpha }\nu
_{\alpha R}\nu _{\alpha R}+f_{\alpha \beta }\nu _{\alpha R}\nu _{\beta R}S\,
\label{ss}
\end{equation}%
where $m_{Di\alpha }$ is the Dirac mass. 
The Majorana masses of the right-handed neutrinos
$M_{\alpha}$ are assumed to be real and diagonal which is
achieved without loss of generality through the choice of our basis.
This Lagrangian has a discrete symmetry, $(-1)^{L}$, where $L$ is the lepton
number, so that $\ell _{i}$ and $\nu _{\alpha R}$ are odd under this
symmetry.

We take the Dirac neutrino masses to be of the order of $0.1$~eV, while the
Majorana masses of the right-handed, singlet, neutrinos are assumed to be of
the order of $0.001$~eV giving rise to pseudo-Dirac neutrinos with very
small mass differences. In order to be consistent with the solar neutrino
data, which does not allow very small mass differences[10], however, two of
the neutrinos are allowed to be only either Dirac or Majorana, but not
pseudo-Dirac. The Dirac case will correspond to three left-handed and three
right-handed neutrinos with the mass matrix given by 
\begin{eqnarray}
M_{\nu } &=&\left( 
\begin{array}{cc}
0 & m_{D}\cr m_{D} & M_{N}%
\end{array}%
\right)  \notag \\
m_{D} &=&\left( 
\begin{array}{ccc}
m_{1} & 0 & 0\cr0 & m_{2} & 0\cr0 & 0 & m_{3}%
\end{array}%
\right) ~~~~  \notag \\
~M_{N} &=&\left( 
\begin{array}{ccc}
0 & 0 & 0\cr0 & 0 & 0\cr0 & 0 & M%
\end{array}%
\right) \,.
\end{eqnarray}%
The two physical neutrino eigenstates ($\nu _{1}$ and $\nu _{2}$) are the
Dirac neutrinos responsible for the solar neutrino oscillations, and only
the state $\nu _{3}$ is a pseudo-Dirac neutrino.

We consider here the second, Majorana, possibility where there is only one
sterile neutrino, which we assume forms a condensate. The corresponding
neutrino mass matrix will be 
\begin{equation*}
M_{\nu }=\left( 
\begin{array}{cccc}
m_{1} & 0 & 0 & 0\cr0 & m_{2} & 0 & 0\cr0 & 0 & 0 & m_{3}\cr0 & 0 & m_{3} & M%
\end{array}%
\right)
\end{equation*}%
Here $\nu _{1}$ and $\nu _{2}$ are Majorana neutrinos, needed to explain the
solar neutrino data, while $\nu _{3}$ is now a pseudo-Dirac neutrino. For
the rest of our analysis, we will restrict our discussions to only the third
eigenstate, the pseudo-Dirac neutrino with a mass matrix 
\begin{equation*}
M_{\nu }=\left( 
\begin{array}{cc}
0 & m_{3}\cr m_{3} & M%
\end{array}%
\right)
\end{equation*}%
and represent the left-handed and right-handed states as $\nu _{L}$ and $\nu
_{R}$. Below we outline the dynamics behind the condensate formation which
we argue is the source for the dark energy.

The exchange of the real scalar, $S$, is found to give rise to attraction
between the right handed neutrinos $\nu _{R}$. Thus, in the context of
cosmic evolution, as the universe cools down to a temperature below the mass
of the neutrinos, this attractive interaction then causes the right-handed
neutrinos to form condensates, the candidates for the dark energy.

The four-Fermion effective contact interaction generated through $S$
exchange is given by 
\begin{equation}
H_{I}=-\mathcal{C}~(\overline{\nu _{M}}~\nu _{M})~(\overline{\nu _{M}}~\nu
_{M})\,.
\end{equation}%
where the right-handed Majorana neutrinos $\nu _{M}$ have been defined as 
\begin{equation}
\nu _{M}=\lambda \nu _{R}+{\nu ^{c}}_{L}\,,
\end{equation}%
where ${\nu ^{c}}_{\alpha L}$ is the CP conjugate of $\nu _{\alpha R}$ and $%
\lambda $ is the Majorana phase, so that the Majorana neutrino satisfies the
Majorana condition $\nu _{M}^{c}=\lambda ^{\ast }\nu _{M}\,.$A number
operator for Majorana particles can not be defined, but in cosmology one can
define the number density per comoving volume of any Majorana particle, when
the interaction rate of the particle is slower than the expansion rate of
the universe. This allows us to define the chemical potential for the
particles. Thus, for the neutrino condensation formation the decoupling
temperature may be considered as the cut off scale for the neutrinos. Below
this temperature, the four-fermi form will be valid with the coupling
strength $\mathcal{C}$ given by 
\begin{equation}
\mathcal{C}={\frac{f^{2}}{m_{S}^{2}}}
\end{equation}%
where $m_{S}$ is the mass of the scalar field and the generation index of
the coupling constant $f$ has been suppressed . 

Below we describe the formalism for the $\nu _{R}$ condensation formation.
It is convenient to work in the Weyl representation, in which the $\gamma
_{5}$ matrix is diagonal. The left-handed and right-handed neutrinos can be
written as 
\begin{eqnarray}
\nu _{L} &=&\left( 1-\gamma ^{5}\right) \nu =%
\begin{bmatrix}
\psi \\ 
0%
\end{bmatrix}
\notag \\
\nu _{R} &=&\left( 1+\gamma ^{5}\right) \nu =%
\begin{bmatrix}
0 \\ 
\overline{\chi }%
\end{bmatrix}
\notag \\
{\nu ^{c}}_{R} &=&{\nu _{L}}^{c}=%
\begin{bmatrix}
0 \\ 
\overline{\psi }%
\end{bmatrix}
\notag \\
{\nu ^{c}}_{L} &=&{\nu _{R}}^{c}=%
\begin{bmatrix}
\chi \\ 
0%
\end{bmatrix}%
\,
\end{eqnarray}%
so that 
\begin{equation*}
\nu _{M}=%
\begin{bmatrix}
\chi \\ 
\lambda \overline{\chi }%
\end{bmatrix}%
;~~\nu _{M}^{c}=%
\begin{bmatrix}
\lambda ^{\ast }\chi \\ 
\overline{\chi }%
\end{bmatrix}%
;~~\overline{\nu _{M}}^{T}=%
\begin{bmatrix}
\lambda ^{\ast }\chi \\ 
\overline{\chi }^{\dag }%
\end{bmatrix}%
;
\end{equation*}

The four-fermion Hamiltonian can then be expressed in terms of the component
fields as 
\begin{eqnarray}
H_{I} &=&-\mathcal{C}\left[ {\lambda ^{\ast }}^{2}~\overline{\chi }%
_{a}^{\dagger }~\chi _{a}~\overline{\chi }_{b}^{\dagger }~\chi _{b}+%
\overline{\chi }_{a}^{\dagger }~\chi _{a}~{\chi _{b}^{\dagger }}~\overline{%
\chi }_{b}\right.  \notag \\
&+&{\chi _{a}^{\dagger }}~\overline{\chi }_{a}~\overline{\chi }_{b}^{\dagger
}~\chi _{b}+\lambda ^{2}~{\chi _{a}^{\dagger }}~\overline{\chi }_{a}~{\chi
_{b}^{\dagger }}~\overline{\chi }_{b}\left. {}\right] \,,
\end{eqnarray}%
and the condensate will correspond to a spin-0 pairing: 
\begin{equation}
\langle \chi _{a}~\overline{\chi }_{b}^{\dagger }\rangle =\epsilon _{ab}~D\,,
\label{a2}
\end{equation}%
giving us the interaction Hamiltonian in the mean-field approximation as: 
\begin{equation}
H_{1}^{MF}=-2~\mathcal{C}\left[ {\lambda ^{\ast }}^{2}~\overline{\chi }%
_{a}^{\dagger }~{\chi }_{b}~D+\lambda ^{2}~{\chi _{a}^{\dagger }}~\overline{%
\chi }_{b}~D^{\ast }\right] ~\epsilon _{ab}\,.
\end{equation}%
In terms of the creation and annihilation operators of any Majorana field: 
\begin{equation}
\psi _{M}(x)=\sum_{p,s}\sqrt{\frac{M_{\alpha }}{2\epsilon }}\left(
f_{ps}u_{ps}e^{-ipx}+\lambda ^{\ast }f_{ps}^{\dagger }v_{ps}e^{ipx}\right)
\,,
\end{equation}%
where $\epsilon ^{2}=p^{2}+m_{i\alpha }^{2}$ is square of the energy of the
physical pseudo-Dirac neutrinos. The interaction Hamiltonian can now be
written as: 
\begin{eqnarray}
H_{1}^{MF} &=&-\mathcal{C}\sum_{p}{\frac{M_{\alpha }}{\epsilon }}  \notag \\
&&\left[ D~{\lambda ^{\ast }}^{2}~e^{-2i\epsilon t}\left( f_{p\uparrow }^{%
\phantom{\dagger}}f_{-p\downarrow }-f_{p\downarrow }f_{-p\uparrow }\right)
\right.  \notag \\
&&+D^{\ast }~{\lambda }^{2}~e^{2i\epsilon t}\left. \left( f_{p\uparrow
}^{\dagger }f_{-p\downarrow }^{\dagger }-f_{p\downarrow }^{\dagger
}f_{-p\uparrow }^{\dagger }\right) \right] \,.  \notag \\
&&
\end{eqnarray}

The complete Hamiltonian will be a sum of the free Hamiltonian ($H_{0}$) and
the interaction Hamiltonian ($H_{1}^{MF}$).It can be transformed to a
canonical form 
\begin{equation}
\mathcal{H}=\sum_{p}E~\left( b_{p\uparrow }^{\dagger }b_{p\uparrow
}+b_{p\downarrow }^{\dagger }b_{p\downarrow }\right) \,,
\end{equation}%
by a time-dependent transformation. Following the standard condensed matter
formalism one finds that a consistent solution, that allows a nonvanishing
condensate $D\neq 0$, gives 
\begin{equation}
{\frac{\mathcal{C}}{2}}\int {\frac{d^{3}p}{(2\pi )^{3}}}{\frac{M_{\alpha
}^{2}}{\epsilon ^{2}}}{\frac{1}{[(\epsilon -\mu )^{2}+\kappa ^{2}]^{1/2}}}%
=1\,,
\end{equation}%
where $\kappa $ is related to the gap parameter, $\Delta $ and $\mu $ is the
chemical potential. The lower limit on the energy integral is taken to be $%
(M_{\alpha }-\mu ),$ and the upper limit is the cut off scale $\Lambda $,
which is the decoupling scale for the neutrinos. This limit is determined by
the requirement that above this temperature neutrino interactions are in
equilibrium and the number density and the chemical potential are not
defined.

\bigskip The solution of (13) gives us the magnitude of the gap: 
\begin{equation}
\Delta =2\sqrt{\frac{2\Lambda }{M_{\alpha }}}\left( 3\pi ^{2}n_{\nu }\right)
^{1/3}e^{-x}\,,
\end{equation}%
where $x=2\pi ^{2}/[\mathcal{C}M_{\alpha }(3\pi ^{2}n_{\nu })^{1/3}]$ and
the critical temperature and the Pippard coherent length are given by 
\begin{eqnarray}
T_{c} &=&{\frac{e^{\gamma }}{\pi }}\approx 0.57\Delta \\
\xi &=&{\frac{e^{x}}{\pi \sqrt{2\Lambda M_{\alpha }}}}\,.
\end{eqnarray}

 In the present example, the coupling of the neutrinos to $S$ becomes strong
in the non-relativistic limit, which will imply that the condition  $M_\alpha\ll m_S$ 
is required to be satisfied for equation (3) to remain valid. However, this condition
may be relaxed considerably in a relativistic treatment of superconductivity. 
A numerical study of fermions interacting
with a scalar field in a strong coupling regime i.e. $f \sim 1$, shows that 
the scalar-field propagator can be scaled as $1/m_S^2$ even for $M_\alpha\geq m_S$ [11]. 
In this strong coupling regime, the right-handed neutrino condensates thus formed start
dominating the universe when the size of Cooper pair becomes comparable to the inter-particle
spacing. For $\xi=0.1$~cm $\sim (2\times 10^{-4}\text{eV})^{-1}$ and $n_{\nu}\sim 110$, 
we get $x\sim 13.5$, $m_S\sim 4.6 \times 10^{-4}$~eV and $\Delta \sim 4\times 10^{-5}$~eV.
The existence of a finite, non zero, gap provides the evidence for a condensate
\footnote{It is interesting to note what happens if we consider the neutrinos in
isolation and discuss their bound state in a non-relativistic framework.
Because of the miniscule mass of the scalar, S, one can approximate the
interaction to be Coulomb-like . Since $f\sim 1$, the Bohr radius of the resulting
atom will be $\sim M_{\alpha}^{-1}$ and the binding energy $\sim M_{\alpha}$ which is essentially of
the order of the dark energy}

This condensate, we would like to argue, is a dark energy candidate. Since
the Cooper pairs are formed around the scale of neutrino masses, the amount
of dark energy density becomes comparable to the matter density in this
scenario. The amount of dark energy is determined by the Majorana mass of
the neutrinos, which is $M\sim 10^{-3}$ eV, and we get the correct order of
magnitude.Thus, without invoking any dynamical field like the quintessence
or accelerons, we have found a natural explanation as to why the scales of
dark energy and neutrino masses are comparable and why dark energy dominates
in this epoch. Furthermore, our model is also consistent with the fact that
the amount of dark energy is the same as the matter density.

We conclude with a short discussion of the basic dynamics of the
condensates, which we call $\xi _{s}.$The Lagrangian for $\xi _{s\text{ }}$%
is given by 
\begin{equation}
{\mathcal{L}}=\left( \partial _{0}+i\mu _{s}\right) \xi _{s}^{\dagger
}\left( \partial _{0}-i\mu _{s}\right) \xi _{s}\,-\partial _{i}\xi
_{s}^{\dagger }\partial _{i}\xi _{s}-V(\xi _{s}),  \label{cond}
\end{equation}%
\noindent where $m $ and $\mu _{s}$ represent the
mass and the chemical potential of the condensate respectively, and 
\begin{equation*}
V(\xi _{s})=m^{2}|\xi _{s}|^{2}+g|\xi _{s}|^{4}\,.
\end{equation*}%
For the present case we assume, $m \simeq 2 m_\nu$ and $\mu_s \simeq \mu$. 
We note that the interaction lagrangian can be written as 
\begin{equation}
V(\xi_s) = m_s^2 |\xi_s \xi_s^\dagger| + {\cal C} (\nu_R^c \nu_R + hc),
\end{equation}
which gives $V(\xi_s)$ after taking into account the quartic self-interactions.
 
In absence of chemical potential, the equation of state for the scalar-field 
becomes
$$\omega=\frac{p}{\rho}=\frac{\text{KE}-V(\xi _{s})}{\text{KE}+V(\xi _{s})}\,. $$ 
For $\text{KE}\ll V(\xi_s)$, the scalar field then behaves as dark energy
with the desired value $\omega \sim -1$. 
This would imply that $g|\xi_s|^2 \ll 0$ or $g<0 $. 
The coupling constant
$g$ can be calculated from the knowledge of scattering between the 
condensates from an analogy with atomic physics. One can write $g=\frac{4\pi a}{m}$ 
for an attractive interaction with $a<0$. Under this condition the condensates can
represent the dark energy [12]. Another possibility is the case when there 
is a finite chemical
potential and there is spontaneous symmetry breaking [8, 13]. 
In this case one can use the arguments
of Ref. [8] to estimate the parameters $m$ and $g$ as
\begin{eqnarray}
m^{2}&=&m_{S}^{2}-\frac{N_{f}\mathcal{C}}{8\pi ^{2}}\left( \Lambda ^{2}-\mu
^{2}\right) \nonumber \\
~~\mathrm{and}~~~g&=&\frac{N_{f}\mathcal{C}}{8\pi ^{2}}ln\left( 
\frac{\Lambda ^{2}}{\mu ^{2}}\right). \,
\end{eqnarray}
In this case the equation of state can be written as 
$\rho=3\,p+4(1-\frac{\mu}{m}\sqrt{1+\sqrt{p}})\sqrt{p}$ [14] which  gives $\omega \sim -1$ if
$p=(\frac{\mu^2}{m^2}-1)^2$ and $\mu^2 >m^2$ which is consistent with the condensate formation
condition [13]. $p$ and $\rho$ are made dimensionless by factoring out with $m^4/4g$.

 In summary we have proposed a new solution of the dark energy problem where
the dark energy is the condensate formed by self interaction of right handed
(singlet) neutrinos generated through the exchange of a singlet scalar.
Since this neutrino is the Dirac partner of one of the three observed
neutrinos, the dark energy is related to the neutrino mass. The fact that
the matter and dark energy are comparable follows naturally from our model.
\subsection*{REFERENCES}

\noindent \lbrack 1] C. Wetterich, Nucl. Phys. B \textbf{302}, 668 (1988);
P.J.E. Peebles and B. Ratra, Astrophys. J. \textbf{325}, L17 (1988).

\noindent \lbrack 2] C.T. Hill, D.N. Schramm, J.N. Fry, Nucl. Part. Phys. 
\textbf{19}, 25 (1989); J.A. Frieman, C.T. Hill, R. Watkins, Phys. Rev. 
\textbf{D 46}, 1226 (1992); A.K. Gupta, C.T. Hill, R. Holman, E.W. Kolb, \ 

Phys. Rev. \textbf{D 45}, 441 (1992); E. Masso, F. Rota, G. Zsembinszki,
Phys. Rev. \textbf{D 70}, 115009 (2004); E. Masso, G. Zsembinszki, JCAP 
\textbf{0602}, 012 (2006); P.Q. Hung, E. Masso, G.

Zsembinszki, JCAP \textbf{0612}, 004 (2006); C.T. Hill, I. Mocioiu, E.A.
Paschos, and U. Sarkar, Phys. Lett. B \textbf{651}, 188 (2007); P.H. Gu,
H.J. He, and U. Sarkar, Phys. Lett. B \textbf{653}, 419

(2007); JCAP \textbf{0711}, 016 (2007); P.H. Gu, arXiv:0710.1044 [hep-ph].

\noindent \lbrack 3] P. Gu, X. Wang, and X. Zhang, Phys. Rev. D \textbf{68},
087301 (2003); R. Fardon, A.E. Nelson, and N. Weiner, JCAP \textbf{0410},
005 (2004); P.Q. Hung, hep-ph/0010126.

\noindent \lbrack 4] H. Li, Z. Dai, and X. Zhang, Phys. Rev. D \textbf{71},
113003 (2005); V. Barger, P. Huber, and D. Marfatia Phys. Rev. Lett. \textbf{%
95}, 211802 (2005); A.W. Brookfield, C. van de Bruck, D.F. Mota,
and D. Tocchini-Valentini, Phys. Rev. Lett. \textbf{96}, 061301 (2006); A.
Ringwald and L. Schrempp, JCAP \textbf{0610}, 012 (2006); R. Barbieri, L.J.
Hall, S.J. Oliver, and A. Strumia, Phys. Lett. B

\textbf{625}, 189 (2005); R. Takahashi and M. Tanimoto, Phys. Lett. B 
\textbf{633}, 675 (2006); R. Fardon, A.E. Nelson, and N. Weiner, JHEP 
\textbf{0603}, 042 (2006); E. Ma and U. Sarkar, Phys. Lett. B

\textbf{638}, 356 (2006); N. Afshordi, M. Zaldarriaga, and K. Kohri,
Phys.Rev. D \textbf{72}, 065024 (2005); O.E. Bjaelde, \textit{et. al.},
JCAP {\bf 0801}:026 (2008),
arXiv:0705.2018v2[astro-ph]; C. Wetterich, Phys. Lett. \textbf{B 655}, 201
(2007); D.F. Mota, V. Pettorino, G. Robbers and C. Wetterich, Phys. Lett. \textbf{B 663}, 160 (2008),
arXiv:0802.1515v1[astro-ph].

\noindent \lbrack 5] J.I. Kapusta, Phys. Rev. Lett. \textbf{93}, 251801
(2004).

\noindent \lbrack 6] S. Antusch, J. Kersten, M. Lindner and M. Ratz, Nucl.
Phys. \textbf{B 658}, 203 (2003).

\noindent \lbrack 7] J.R. Bhatt and U. Sarkar, Phys. Rev. \textbf{D 80}, 045016 (2009);
arXiv:0805.2482[hep-ph].

\noindent \lbrack 8] G. Barenboim, JHEP \textbf{ 0903}: 102 (2009);
arXiv:0811.2998[hep-ph].

\noindent \lbrack 9] P. Minkowski, Phys. Lett. \textbf{67B}, 421 (1977); T.
Yanagida, in \textit{Proc. of the Workshop on Unified Theory and the Baryon
Number of the Universe}, ed. O. Sawada and A. Sugamoto
(KEK, Tsukuba, 1979), p. 95; M. Gell-Mann, P. Ramond, and R. Slansky, in 
\textit{Supergravity}, ed. F. van Nieuwenhuizen and D. Freedman (North
Holland, Amsterdam, 1979), p. 315; S.L.
Glashow, in \textit{Quarks and Leptons}, ed. M. L$\mathrm{\acute{e}}$vy 
\textit{et al.} (Plenum, New York, 1980), p. 707; R.N. Mohapatra and G.
Senjanovi$\mathrm{\acute{c}}$, Phys. Rev. Lett. \textbf{44}, 912 (1980).

\noindent \lbrack 10] A.de Gouvea, W.-C. Huang and J. Jenkins, Phys.Rev. \textbf {D 80},073007 (2009);
 arXiv:0906.1611[hep-ph].

\noindent \lbrack 11] R.D. Pissarski and D.H. Rischke, Phys.Rev.  \textbf{D
60}, 094013 (1999).

\noindent \lbrack 12] T. Fukuyama, M. Morikawa and T. Tatekawa, JCAP {\bf 0806}:033 (2008);
arXiv:0705.3091[astro-ph]

\noindent \lbrack 13] A. H. Rezaein and H. J. Pirner, Nucl. Phys. \textbf{A 779}, 197 (2006);
arXiv:0606043 [nucl-th].

\noindent \lbrack 14] J.R. Bhatt and V. Sreekanth; arXiv:0910.1972 [hep-ph].
\end{document}